\documentclass[useAMS,usenatbib]{mn2e}

\usepackage{epsfig}
\usepackage{amsmath}
\usepackage{amssymb}
\usepackage{natbib}

\usepackage{epstopdf}

\usepackage{multicol}

\newcommand{\swift}{\textit{Swift}}

\newcommand{\xmm}{\textit{XMM-Newton}}

\newcommand{\Msun}{\mathrm{M}_{\odot}}
\newcommand{\lum}{\mathrm{erg~s}^{-1}}
\newcommand{\nh}{\mathrm{cm}^{-2}}

\newcommand{\cnts}{\mathrm{counts~s}^{-1}}

\newcommand{\xte}{XTE J1701--462}
\newcommand{\grs}{GRS J1741--2853}
\newcommand{\ks}{KS 1741--293}

\usepackage[dvips]{color}

\def \mnras {MNRAS}
\def \apj {ApJ}
\def \apjs {ApJS}
\def \apjl {ApJL}
\def \aap {A\&A}

\def \araa {ARAA}

\title[Low-level accretion flare in SAX J1750.8--2900]{A low-level accretion flare during the quiescent state of the neutron-star X-ray transient SAX J1750.8--2900}

\author[Wijnands \& Degenaar]
{R. Wijnands$^{1}$\thanks{e-mail: r.a.d.wijnands@uva.nl},  
N. Degenaar$^{1}$\thanks{Hubble Fellow}\\
$^{1}$Astronomical Institute "Anton Pannekoek", 
University of Amsterdam, 
Postbus 94249, 1090 GE Amsterdam, The Netherlands\\
$^{2}$Department of Astronomy, University of Michigan, 500 Church Street, Ann Arbor, MI 48109-1042, USA\\
}

\begin{document}


\pagerange{\pageref{firstpage}--\pageref{lastpage}} \pubyear{0000}

\maketitle

\label{firstpage}

\begin{abstract} We report on a series of \swift/XRT observations,
performed between February and 22 March 2012, during the quiescent
state of the neutron-star X-ray binary SAX J1750.8--2900. In these
observations, the source was either just detected or undetected,
depending on the exposure length (which ranged from $\sim$0.3 to
$\sim$3.8 ks). The upper limits for the non-detections were consistent
with the detected luminosities (when fitting a thermal model to the
spectrum) of $\sim10^{34}$ erg s$^{-1}$ (0.5--10 keV). This level is
consistent with what has been measured previously for this
source in quiescence. However, on March 17 the source was found to
have an order of magnitude larger count rate. When fitting the flare
spectrum with an absorbed power-law model, we obtained a flare
luminosity of $(3-4) \times 10^{34}$ erg s$^{-1}$ (0.5--10 keV).
Follow-up \swift\ observations showed that this flare lasted
$<16$~days. This event was very likely due to a brief episode of
low-level accretion onto the neutron star and provides further
evidence that the quiescent state of neutron-star X-ray transients
might not be as quiet as is generally assumed.  The detection of this
low-level accretion flare raises the question whether the quiescent
emission of the source (outside the flare) could also be due to
residual accretion, albeit continuous instead of episodic. However, we
provide arguments which would suggest that the lowest intensity level
might instead represent the cooling of the accretion-heated neutron
star.

\end{abstract}


\section{Introduction} 

In neutron-star low-mass X-ray binaries (LMXBs) a neutron star is
accreting mass from a donor with a mass lower (often much lower) than
the mass of the neutron star. The mass transfer occurs because the
donor star fills its Roche lobe. Most systems (the X-ray transients)
are only occasionally visible during short X-ray outbursts (with X-ray
luminosities from $\sim 10^{34}$ erg s$^{-1}$ up to a few times
$10^{38}$ erg s$^{-1}$) but most of the time the sources are in their
dormant, quiescent state (during which they have X-ray luminosities of
typically $10^{32-34}$ erg s$^{-1}$). In this state, many systems have
a X-ray spectrum that is dominated by a soft thermal component with a typical black-body temperature of $\sim$0.1--0.3 keV. Often a
non-thermal component is present as well which can be described by a
power-law with photon index of $\sim$1--2 \citep[see, e.g., ][]{1998A&ARv...8..279C}. In some sources this
non-thermal component dominates the 0.5--10 keV X-ray spectrum, and no
thermal emission can significantly be detected
\citep[e.g.,][]{campana2005_amxps,wijnands2005,wijnands05_amxps,heinke2009,degenaar2012_amxp}.

The origin of the non-thermal component is not understood and it has
been speculated that the neutron star magnetic field plays an
important role in the production of this spectral component
\citep[see, e.g.,
][for reviews]{1998A&ARv...8..279C,degenaar2012_amxp}. The
thermal component very likely arises from the surface of the neutron
star. This component could represent the cooling emission of the
neutron star core which is heated in outburst due to pycno-nuclear
reactions deep in the neutron star crust \citep[][]{brown1998}. If true, then the
quiescent thermal flux (which is related to the core temperature) and
the long-term time averaged accretion rate (representing the amount of
heating in the crust) should be positively correlated. It has been
found that most systems have cores which are too cold to be explained
by standard core cooling processes and enhanced neutrino emission
processes are needed in the core to cool them down \cite[e.g.,][]{brown1998,yakovlev2004,heinke2009,wijnands2012}; only a few
systems can be described accurately with standard core cooling
\citep[e.g.,][]{rutledge1999,rutledge2000,degenaar2010_exo2,wijnands2012}. Enhanced
core cooling is thought to occur for neutron stars that are relatively
massive or are composed of exotic matter
\citep[e.g.,][]{2001ApJ...548L.175C,lattimer2001,page2004}.

Alternatively, the soft thermal component could be due to low-level
residual accretion down to the neutron star surface. When the matter
falls onto the neutron star, it emits radiation with a thermal-like
spectrum which is nearly indistinguishable (i.e., when taking into
account the often very low number of photons detected) from the thermal
spectrum expected from the cooling of the neutron stars
\citep[][]{zampieri1995,soria2011}. If this scenario occurs, those
systems would be excellent targets to search for gravitational
red-shifted lines due to heavy elements which are accreted onto the
neutron star surface \citep[][]{bildsten1992,brown1998}. If no
accretion occurs, those elements sink very quickly into the crust and
only hydrogen remains at the top which does not have lines in the soft
X-ray band.

Clearly, it is important to determine weather the thermal quiescent
emission is due to residual accretion or due to core cooling. However,
this is not very straightforward to determine.  One way to distinguish
residual accretion from cooling as the dominant source of the thermal
emission is to search for variability of the quiescent emission since
core cooling is expected to be much more stable in time than the
expected stochastic variations in the residual accretion
rate. Therefore, variability of the thermal component would be clear
evidence that residual accretion might be occurring in the system. We
note that crust cooling from a heated neutron-star crust due to long
episodes of accretion could also produce detectable variability in
quiescence
\citep[e.g.,][]{wijnands2002,wijnands2004,cackett2008,cackett2010,degenaar2010_exo2,fridriksson2011},
but typically this is expected to be a smooth decay
\citep[e.g.,][]{rutledge2002,shternin07,brown08,page2012}.

Here we report clear evidence for a short, very-faint accretion
episode in the neutron-star X-ray transient SAX J1750.8--2900 during
its quiescent phase using observations performed with the X-ray
telescope (XRT) aboard \swift. SAX J1750.8--2900 was discovered by
\citet{natalucci1999} using the {\it BeppoSAX} satellite. The
detection of type-I X-ray burst demonstrated the presence of a
neutron-star accretor and allows for a distance estimate of $\sim$6.8
kpc \citep[][]{galloway06}. After its discovery outburst, the source
exhibited several more outbursts
\citep[e.g.,][]{kaaret2002,markwardt2008,linares2008}, with the last,
relatively faint (with a peak luminosity of $\sim 10^{36}$ erg
s$^{-1}$), outburst occurring in February 2011
\citep[][]{fiocchi2011,natalucci2011}. It is unclear how long
this 2011 outburst lasted, but the source light curve obtained through the Galactic bulge scan program\footnote{The RXTE/PCA
Galactic bulge scan light curves  \citep[][]{2001ASPC..251...94S} can be found at:
http://asd.gsfc.nasa.gov/Craig.Markwardt//galscan/main.html} \citep[][]{2001ASPC..251...94S} using the Proportional Counter Array (PCA) aboard the Rossi X-ray Timing Explorer (RXTE) satellite suggests
that the outburst lasted at most until the end of February/early March
2011. No outbursts have been reported from the source since this 2011
outburst.

\section{Observations and results}

SAX J1750.8--2900 was observed with {\it Swift}/XRT between 2012
February 14 and March 17 during a series of observations spaced by
1--2 weeks (see Table~\ref{tab1} for a log of the observations). All
data were obtained in photon counting (PC) mode. Typically the source
intensity was near the detection limit, with the source either being
undetected or showing a small excess of photons (depending on exposure
time; see Table~\ref{tab1}). On March 17, however, we noticed a large
enhancement in the source count rate. We obtained two extra
observations within the following week to study the decay of this
flare.

We used the {\it Swift} data analysis tools incorporated in HEASOFT
version 6.11. We processed all data using the tool {\em
xrtpipeline}. We used the tool {\em Xselect} to extract images, light
curves and spectra. A circular region with a radius of 10 pixels was
used to extract source events and three circular regions with radii of
10 pixels to extract the background data. We created exposure maps for
each observation and used these when creating light curves and
spectra.

The source was detected (albeit sometimes barely) when the
observations were sufficiently long ($>$ 3 ks) and during the two
flare observations (Obs IDs 31174030--31 in Table~\ref{tab1}). For the
non-detections during the shorter observations we determined upper
limits on the count rate (95\% confidence level) using the
prescription of \citet{gehrels1986}. We also co-added all pre-flare
data (i.e., excluding the three last observations listed in
Table~\ref{tab1}, which concern the observation in which the flare was
discovered and the two follow-up observations): the source is clearly
detected in this combined data set at a count rate of $(9.1\pm2.8)\times10^{-4}~\cnts$.

\begin{figure}
 \begin{center}
 \includegraphics[width=8cm]{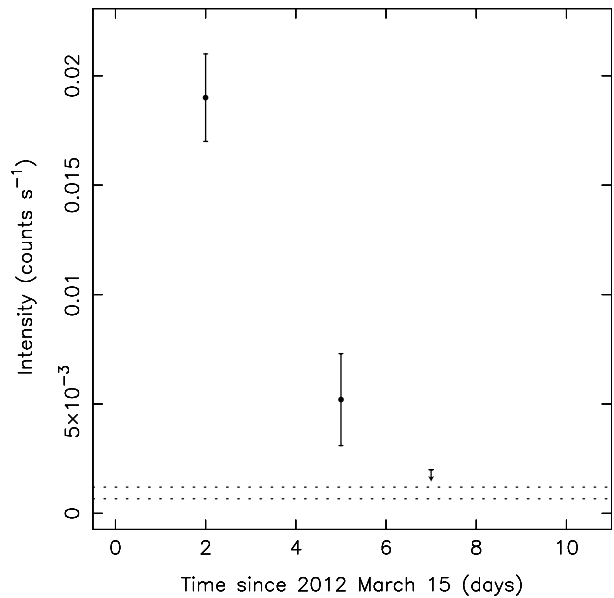}
    \end{center}
 \caption[]{\swift/XRT count rate curve (0.3--10 keV) of the flare episode. The dotted lines indicate the quiescent count rate observed during the pre-flare observations. The count rate upper limit obtained during the last observation is for 95\% confidence level.}
 \label{figlc}
\end{figure}

The light curve of the flare episode is shown in
Figure~\ref{figlc}. The source clearly exhibited a flare on March 17,
with a count rate a factor of $\sim10$ above the quiescent level (see
also Table~\ref{tab1}). The data obtained on March 20 shows that the
source intensity was still enhanced compared to the pre-flare level by
a factor of $\sim5$, but on March 22 the count rate was consistent
with the source being back in quiescence. Since the last observation just
before the flare occurred on March 6, we can constrain the flare
duration between 5 and 16 days.

\begin{figure}
 \begin{center}
\includegraphics[width=6cm,angle=-90]{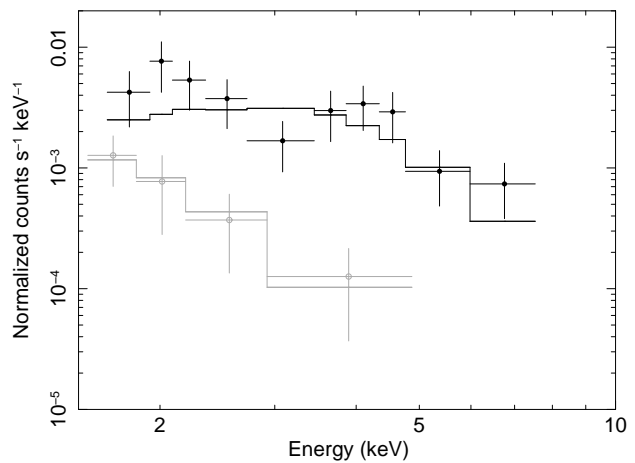}
    \end{center} \caption[]{{\it Swift}/XRT spectra of the peak of the flare (black; Obs IDs 31174030--31) and the combined quiescent data (grey; all other observations; see Table~\ref{tab1}).}
 \label{figspectra}
\end{figure}

We extracted the spectra of the flare observations (Obs IDs
31174030--31; Table~\ref{tab1}), as well as the combined quiescent
data set (all observations except Obs IDs 31174030--31). The obtained
spectra were grouped to a minimum of 5 photons per bin. Although this
is not sufficient to formally allow to use the $\chi^2$ minimization
technique to fit the spectra, we have found previously
\citep[e.g.,][]{2002ApJ...568L..93W,2013MNRAS.428.3083A} that this
method can still be used accurately even with so few photons per
spectral bin (we note that when using the W-statistics, we
obtained consistent results). We created ancillary response files
(using exposure maps) and used the response matrix files (v.13) from
the CALDB database. The spectra were fit in the range 0.5--10 keV
using \textit{XSpec}. The errors on the fit parameters are for 90\%
confidence levels. The quiescent and flare spectra are shown in Figure
\ref{figspectra}.

Despite that the source was clearly detected both during the flare and
in the combined quiescent data set, the extracted spectral data were
of insufficient quality to fit complex models. We therefore fixed the
column density $N_{\rm H}$ in order to investigate the spectral
shape. The $N_{\rm H}$ for SAX J1750.8--2900 is not well constrained,
so here we use values of $4\times 10^{22}$ cm$^{-1}$ and $6\times
10^{22}$ cm$^{-1}$ which were obtained by analyzing \xmm\ quiescent
spectral data of the source \citep[][]{lowell2012}. The assumed column
density did not significantly impact the obtained spectral parameters
(see Table~\ref{tab2}).

The flare spectrum could be fitted with an absorbed power-law model
(PHABS*POWERLAW), yielding a reduced chi-squared value of
$\chi^2_{\nu}=1.3$ for 8 degrees of freedom (dof). The obtained photon
index was relatively high (2.4--2.8) albeit with large errors
(1.3--1.6). The inferred flux during the flare was $\sim 10^{-12}$ erg
s$^{-1}$, and the unabsorbed flux was $\sim (5 - 8) \times 10^{-12}$
erg s$^{-1}$ (both 0.5--10 keV). The large range in the latter is due
to the range in assumed $N_{\rm H}$. This demonstrates that when the
absorption is high and not well constrained, extrapolating a power-law
model to low energies ($<2$ keV) results in large uncertainties in the
unabsorbed fluxes. Using these fluxes, we obtain a source luminosity
during the flare of $\sim (3 - 4) \times 10^{34}$ erg s$^{-1}$ for a
distance of 6.8 kpc.

A power-law model could also describe the quiescent spectrum
satisfactory but the resulting photon index was very large ($>$6),
suggesting that a thermal model is more appropriate. To allow for a
direct comparison with the quiescent spectrum observed with \xmm, we
follow \citet{lowell2012} and fit the XRT spectrum with a black-body
model (BBODYRAD) with $N_{\rm H} = 4\times 10^{22}~\nh$ fixed, and a
neutron star atmosphere model \citep[NSATMOS;][]{heinke2006} with
$N_{\rm H} = 6\times 10^{22}~\nh$ fixed. Both give acceptable fits
with $\chi^2_{\nu}=0.5$ (2 dof) and $1.0$ (3 dof) for the black body
and NSATMOS model, respectively.\footnote{For the NSATMOS model fits
the source distance was fixed at 6.8 kpc and the neutron star mass and
radius at $1.4~\Msun$ and $10$~km, respectively. We also fixed
the fraction of the neutron star that is emitting to one (the whole
surface is emitting). For the black-body fit the normalization (which
equals the radius in km squared divided by the distance in units of 10
kpc squared) is poorly constrained ($19^{+4777}_{-19}$).} The obtained
black-body temperature is $0.3\pm0.2$ keV, and the temperature for the
NSATMOS model is $0.15\pm0.02$ keV. For both assumed column densities
and both spectral models the 0.5--10 keV absorbed flux is $\sim 6
\times 10^{-14}$ erg s$^{-1}$, and the unabsorbed flux $\sim(1.2-1.6)
\times 10^{-12}$ erg s$^{-1}$. The resulting 0.5--10 keV luminosities
are $ (6 - 9) \times 10^{33}$ erg s$^{-1}$ (for a distance of 6.8
kpc). Our obtained quiescent temperatures and fluxes are similar to
that found for SAX J1750.8--2900 by \citet{lowell2012} using \xmm\
data. We have also tried to fit the flare spectrum with both the
BBODYRAD and the NSATMOS model, but the spectrum could not be
adequately fit with those models (with reduced $\chi^2$ ranging from
1.8 to 2.3). This clearly demonstrates that the flare spectrum cannot
be described by a single thermal component and therefore it is quite
different than the quiescent spectrum. We note that multiple component
models, e.g., a power law component plus a thermal component, could
fit the data adequately but no meaningful constraints on the thermal
component could be obtained due to the low statistical quality of our
data.

\section{Discussion}

The neutron-star X-ray transient SAX J1750.8--2900 was monitored in
quiescence using {\it Swift}/XRT. During one of those observations, we
detected a faint flare which lasted $<16$ days (as determined using
triggered follow-up observations with {\it Swift}/XRT) and had an
estimated 0.5--10 keV peak luminosity of $\sim (3-4) \times 10^{34}$
erg s$^{-1}$, which is a factor of $\sim$3--4 times higher than the
quiescent level observed with the XRT. Our observed quiescent
luminosity is consistent with that measured previously by
\citet{lowell2012} using \xmm.

Similar flares reaching a factor $\sim10$ above the quiescent level
have been seen in a few transient neutron star LMXBs. Regular
monitoring observations of the Galactic center with \swift/XRT
\citep[][]{degenaar09_gc,degenaar2010_gc} caught an accretion flare
from \ks\ that reached a 2--10 keV peak luminosity of $\sim
3\times10^{34}~\lum$ and lasted $<4$ days
\citep[][]{degenaar2013_ks1741}. That same monitoring program also
detected a short flare from \grs, which reached up to $\sim
9\times10^{34}~\lum$ (2--10 keV) and had a duration of $\sim1$ week
\citep[][]{degenaar09_gc}. In both sources, the short-lived flare was
followed by a brighter and longer (i.e., normal) outburst within a few
weeks (both sources are very regularly active). For SAX J1750.8--2900
we did not observe a new outburst shortly after the flare, although
faint outbursts (with peak luminosities below $10^{36}$ erg s$^{-1}$)
are easily missed. Another source, \xte, displayed several weak flares
(with a peak luminosity of $\sim 10^{34-35}~\lum$; 2--10 keV), with
the first one occurring within 1 year after the end of its most recent
outburst, but others occurred several years ($>$ 3 years) later
\citep[][]{fridriksson2010,fridriksson2011}.

\subsection{Accretion flares}

A natural explanation for these faint flares appears to be low-level
accretion activity during quiescence, although such events are not
easy to understand within the disk instability model that provides the
frame work to describe the observed outburst--quiescent cycles of
transient neutron star LMXBs \citep[e.g.,][]{lasota01}. The processes
at work at the very low accretion luminosities we observed during the
flare of SAX J1750.8--2900 and those we observe in the other sources
are not well understood.  When fitting the flare spectrum of SAX
J1750.8--2900 with an absorbed power-law model, the obtained photon
index is rather high (although with large error bars), which is
similar to what has been seen for other neutron-star X-ray binaries at
such luminosities
\citep[e.g.,][]{2005A&A...440..287I,2009A&A...506..857I,2011MNRAS.417..659A,armas2013,2013ApJ...767L..31D}. This
relatively soft spectrum is consistent with the accretion models which
assume that at those luminosities the flow can be described as a
radiation inefficient accretion flow or as an advection dominated
accretion flow \citep[see, e.g., the discussions
in][]{2004ApJ...601..439T,2011MNRAS.417..659A}. 

However, when high quality data are obtained at those luminosities,
the situation becomes more complex: a prominent thermal component can
clearly be detected \citep[][]{2013ApJ...767L..31D,armas2013}, most
probably due to thermal emission from the neutron star surface. This
thermal emission is likely caused by low-level accretion onto the
surface.  When including this soft component in the spectral models
the power-law component becomes significantly harder (with photon
indices of $\sim$1.5) than when only a single power-law component is
fitted to the data. This might indicated that at different
luminosities the power law component is caused by different physical
processes \citep[see also the discussion in][]{armas2013}.  Sadly, the
quality of our flare spectrum does not allow us to determine what
exactly causes the relative softness of the flare spectrum of SAX
J1750.8-2900. Clearly, to make progress in understanding the accretion
processes at very low accretion rates, high quality data (i.e.,
obtained with {\it XMM-Newton}) need to be acquired to study the
different spectral components in more detail. However, the short
duration of the flare ($<$16 days; similar to the flare durations seen
in other sources) indicates that such accretion flow can come and go
on relatively short time scales and this needs to be incorporated in
the models.

\subsection{Origin of the quiescent emission}

The fact that we see a flare from SAX J1750.8--2900 that is likely
related to a short and weak accretion event, raises the question
whether low-level accretion is also relevant for the out-of-flare
quiescent emission. If a residual accretion flow would reach the
neutron star surface, it would very likely generate a thermal spectrum
that is indistinguishable from that of a cooling neutron star
\citep[e.g.,][i.e., given the low statistics of quiescent data from
neutron-star X-ray transients]{zampieri1995}. \citet{lowell2012}
discuss the quiescent properties of the source, as observed with \xmm.
They used the cooling curves presented by \citet{yakovlev2004} to
compare the observed quiescent luminosity with that expected from a
cooling neutron star and reach the (correct) conclusion that the
source is too bright (thus too hot) and it cannot be explained using
those cooling curves. This could suggest that low-level accretion onto
the neutron star is indeed important in quiescence. The accretion
luminosity would then mask the cooling emission and the neutron star
would be significantly coolder than inferred and would be consistent
with the cooling models. However, significant uncertainties exist in
the theoretical cooling curves and when using slightly different
assumptions, cooling curves can be constructed that allow such a hot
neutron star: e.g., using the mass accretion rate and the observed
quiescent luminosity for the source as given in \citet{lowell2012}, it
can be seen that the source falls right on top of the Bremsstrahlung
cooling curve presented in Figure 1 of
\citet{wijnands2012}. Therefore, it is feasible that the source indeed
harbors a hot neutron star. It would then be one of the few whose core
cools very slowly due to the absence of strong neutrino emission
processes.

In addition, several extra arguments can be given that in we do
observe the cooling emission from SAX J1750.8--2900 instead of
radiation due to residual accretion.  \citet{lowell2012} do not
observe a power-law component in the quiescent spectrum obtained with
\xmm\ (consistent with the \swift\ results, although our constraints
are not very strong). Although it is not fully clear whether at low
accretion rates indeed a power-law component should be present
\citep[possibly only a thermal component will be seen if the accretion
rate is low enough and the matter falls onto the neutron star
surface;][]{zampieri1995}, during the flare the spectrum was totally
dominated by a power-law despite that the luminosity was only a factor
of 3--4 higher than the quiescent level. One might therefore expect
that at only slightly lower accretion rate, also a power-law component
might be present. Furthermore, our inferred quiescent luminosity is
consistent with that reported by \citet{lowell2012}, which was
measured $\sim2$ years previously (in addition a faint outburst
occurred between our observations and theirs). This strongly indicates
that at this luminosity level the source is not highly variable, as
one would expect it accretion would play a significant role.

\noindent {\bf Acknowledgements.}\\ R.W. acknowledges support from a
European Research Council (ERC) starting grant. N.D. is supported by
NASA through Hubble postdoctoral fellowship grant number
HST-HF-51287.01-A from the Space Telescope Science Institute, which is
operated by the Association of Universities for Research in Astronomy,
Incorporated, under NASA contract NAS5-26555. The authors acknowledge
the use of public data from the {\it Swift} data archive. {\it Swift}
is supported at PSU by NASA contract NAS5-00136. This research has
made use of the XRT data Analysis Software (XRTDAS) developed under
responsibility of the ASI Science Data Center (ASDC), Italy. This
research has made use of NASA's Astrophysics Data System.
 
\bibliographystyle{mn2e}

\begin{table}
\caption{Log of the \swift/XRT observations.}
\begin{tabular}{cccc}
\hline
Observation ID & Date & Exposure time & Count rate \\
               & (2012) & (ks)    & (counts~s$^{-1}$)  \\
\hline
31174024 &	14 Feb	&	3.8	& $(1.5\pm0.6)\times10^{-3}$	 \\ 
31174025 &	26 Feb	&	2.6	& $<1.8\times10^{-3}$\\ 
31174026 &	29 Feb	&	0.3	& $<1.2\times10^{-2}$\\ 
31174027 &	 3 Mar	&	3.2	& $(1.6\pm0.7)\times10^{-3}$	 \\ 
31174028 &	 6 Mar	&	2.8	& $<1.7\times10^{-3}$\\ 
31174030 &	17 Mar	&	3.1	& $(1.9\pm0.2)\times10^{-2}$	 \\ 
31174031 &	20 Mar	&	1.0	& $(5.2\pm2.1)\times10^{-3}$	 \\ 
31174032 &	22 Mar	&	1.0	& $<2.0\times10^{-3}$	\\ 
\hline
\multicolumn{4}{l}{Quoted count rates are background corrected and for the full}\\
\multicolumn{4}{l}{XRT energy range (0.3--10 keV).}
\end{tabular}
\label{tab1}
\end{table}

\begin{table*}
\caption{Spectral fit results}
\begin{tabular} {lllllll}
\hline
Parameter        & $\chi^2_{\nu}$ (dof)     &  Photon index    & Temperature$^a$ & $F_{\rm abs}^b$            & $F_{\rm unabs}^b$ & $L_{\rm X}^c$ \\
                      &         &         &  (keV)                       &  ($10^{-13}$ erg s$^{-1}$) & ($10^{-13}$  erg s$^{-1}$) & ($10^{34}~\lum$) \\
\hline                 
\multicolumn{7}{c}{$N_{\rm H}$ = $4 \times 10^{22}$ cm$^{-2}$}\\
\hline
Flare                 & 1.3 (8) & $2.4\pm1.3$     &                                      &  $12\pm5$                & $46\pm 16$  & $2.5\pm 0.8$ \\
Quiescence             &                  &                        &                 &                          &    & \\
 - Black body    & 0.5 (2)    &                  & $0.3\pm0.2$                          &  $0.6\pm0.2$             & $12\pm5$ & $0.6\pm 2.8$\\
\hline
\multicolumn{7}{c}{$N_{\rm H}$ = $6\times 10^{22}$ cm$^{-2}$}\\
\hline
Flare          & 1.5 (8)       &  $2.8\pm1.6$     &                                &  $11\pm5$                & $78\pm 15$ & $4.2\pm 0.8$\\
Quiescence             &                  &                        &                 &                          &    &  \\
 - NSATMOS       & 1.0 (3)         &                  &        $0.15\pm0.02$   &  $0.6\pm0.2$             & $16\pm7$ & $0.9\pm 0.3$\\
\hline
\multicolumn{7}{l}{$^a$ For the NSATMOS model we assumed a neutron-star mass of 1.4~$\Msun$, a radius of 10 km, and a distance of 6.8 kpc}\\
\multicolumn{7}{l}{$^b$ The fluxes are for the energy range 0.5--10 keV. The errors on the fluxes are determined using the method outlined in \cite{wijnands2004}}\\
\multicolumn{7}{l}{$^c$ The luminosities are calculated from the unabsorbed 0.5--10 keV flux by assuming a distance of 6.8 kpc.}\\
\end{tabular}
\label{tab2}
\end{table*}

\label{lastpage}
\end{document}